\documentclass[11pt,a4paper]{article}
\usepackage[T1]{fontenc}
\usepackage[utf8]{inputenc}
\usepackage{lmodern}
\usepackage[margin=2.3cm]{geometry}
\usepackage{amsmath,amssymb,bm,graphicx,booktabs,array,tabularx}
\usepackage{microtype}
\usepackage[numbers,sort&compress]{natbib}
\usepackage[hidelinks]{hyperref}
\hypersetup{pdftitle={A conservative bulk-surface closure for soluble-surfactant falling films},pdfauthor={Sanghasri Mukhopadhyay and Asim Mukhopadhyay}}
\usepackage{caption}
\newcommand{\Rey}{\mathrm{Re}}
\newcommand{\Pe}{\mathrm{Pe}}
\newcommand{\Ca}{\mathrm{Ca}}
\newcommand{\Ka}{\mathrm{Ka}}
\newcommand{\Mn}{\mathrm{Mn}}
\newcommand{\mass}{\mathcal{M}}
\newcommand{\dd}{\mathrm{d}}
\newcommand{\ReferenceMassChange}{-2.28832050\times10^{-4}}
\newcommand{\CorrectedMassChange}{-1.15108\times10^{-15}}
\newcommand{\ReferenceLossPercent}{0.0228832}
\newcommand{\ReferenceFinalRate}{-4.65796039\times10^{-6}}
\newcommand{\RateIdentityError}{9.2\times10^{-17}}
\newcommand{\TranslationResidual}{1.83\times10^{-3}}
\newcommand{\InstantaneousConservativeRate}{1.24\times10^{-18}}
\newcommand{\DifferenceH}{2.09\times10^{-5}}
\newcommand{\DifferenceQ}{5.05\times10^{-5}}
\newcommand{\DifferenceB}{9.36\times10^{-6}}
\newcommand{\DifferenceGamma}{4.86\times10^{-5}}

\title{A conservative bulk-surface closure for soluble-surfactant falling films: an analytical and numerical mass-balance audit}
\author{Sanghasri Mukhopadhyay\thanks{Corresponding author: \href{mailto:sanghasri.m@iiitb.ac.in}{sanghasri.m@iiitb.ac.in}}\textsuperscript{1}
\quad and\quad Asim Mukhopadhyay\textsuperscript{2}\\[5pt]
\small \textsuperscript{1}Centre for Applied Sciences, International Institute of Information Technology,\\[-2pt]
\small Bengaluru 560100, India\\[3pt]
\small \textsuperscript{2}Department of Mathematics, Vivekananda Mahavidyalaya,\\[-2pt]
\small Burdwan 713103, West Bengal, India}
\date{}
\begin{document}
\maketitle
\begin{abstract}
Adsorption and desorption redistribute soluble surfactant between a liquid film and its free surface, but cannot change the total inventory in a closed periodic domain. We audit this requirement in the nonlinear weighted-residual models of Mukhopadhyay et al.\ [Phys.\ Rev.\ Fluids \textbf{11}, 084004 (2026); Phys.\ Fluids \textbf{38}, 092118 (2026)]. Their correction to the nonlinear surface Marangoni term restores a conservative interfacial flux, but leaves a separate inconsistency in the depth-integrated bulk diffusion. We derive the resulting mass imbalance and construct explicit positive periodic concentration fields that violate exact conservation. The missing contribution arises from the streamwise concentration gradient appearing in the normal diffusive flux at a sloping interface. Including this term ensures local conservation of surfactant across the bulk and free surface, while leaving the hydrodynamic equations, the leading-order adsorption–desorption relation and the linearized equations about uniform equilibrium unchanged. In an independent no-slip calculation using the parameters of the earlier traveling-wave study, the reference model loses $0.0228832\%$ of its initial surfactant mass by $t=4500$, whereas the corrected run has an absolute relative mass change of $1.2\times10^{-15}$ at the same time. The main hump and capillary ripples remain closely comparable at around  $t=4500$. Although the drift is small in this particular case, the corrected equations conserve the reduced total inventory for every smooth periodic solution, so that the closed-system mass balance is built directly into the nonlinear model. We also distinguish this exact conservation property from higher-order compatibility with the full curved-interface boundary conditions.  Therefore, this benchmark shows that agreement with traveling-wave calculations does not, by itself, guarantee that a reduced model respects the conservation laws inherited from the underlying transport problem.  
\end{abstract}
\noindent\textbf{Keywords:} falling films; soluble surfactant; weighted-residual model; mass conservation; interfacial transport.

\section{Introduction}
A soluble surfactant moves through a liquid film and along its free surface. Adsorption and desorption transfer material between these two reservoirs, while gradients in surface concentration generate Marangoni stresses that modify the flow. This coupling makes the treatment of surfactant transport important for nonlinear film dynamics. It also provides a direct check on a reduced model: with an impermeable wall, periodic boundaries and no external source, the total surfactant inventory must remain constant.

The mass balance on a deforming interface accounts for changes in both concentration and surface area. Stone~\cite{Stone1990} derived this balance, and Wong, Rumschitzki and Maldarelli~\cite{Wong1996} clarified its interpretation. For soluble surfactants, the bulk and surface balances must use equal and opposite exchange terms. A long-wave reduction must preserve that cancellation and treat the normal diffusive flux consistently with the changing film thickness.

Weighted-residual models describe nonlinear film dynamics at much lower computational cost than the full free-boundary problem. The approach developed by Ruyer-Quil and Manneville~\cite{RuyerQuil2000} was extended to soluble-surfactant films by Pascal, D'Alessio and Ellaban~\cite{Pascal2019}, and to heated surfactant-laden films by D'Alessio et al.~\cite{DAlessio2020}. These models use approximate velocity and concentration profiles to obtain depth-integrated evolution equations. Agreement with linear stability theory is an essential check, but it does not establish a nonlinear conservation law.

Mukhopadhyay et al.~\cite{Mukhopadhyay2026PRF,Mukhopadhyay2026POF} identified an inconsistency in the nonlinear surface Marangoni term of the earlier models ~\cite{Pascal2019, DAlessio2020}. In a common no-slip notation, the earlier term can be written as
\begin{equation}
 \frac{M}{4}\left(h\Gamma\Gamma_{xx}+h\Gamma_x^2+5\Gamma\Gamma_xh_x\right)
 =\frac{M}{4}\partial_x(h\Gamma\Gamma_x)+M\Gamma\Gamma_xh_x,
 \label{eq:surface-old}
\end{equation}
where $M=\Rey\Mn$; an explicit long-wave scaling supplies an additional factor $\varepsilon$. Replacing the coefficient $5/4$ of $M\Gamma\Gamma_xh_x$ by $1/4$ removes the final nonflux term. We retain that correction throughout this study.

A separate issue remains in the bulk equation. In the models of Refs.~\cite{Mukhopadhyay2026PRF,Mukhopadhyay2026POF}, the depth-integrated diffusion term contains
\begin{equation}
 \frac{1}{\Pe_b}\left(\chi_{xx}+h\varphi_{xx}-\frac{3\chi h_x^2}{h^2}\right),
 \label{eq:bulk-bracket-intro}
\end{equation}
whose integral over a period need not vanish. Cancellation of adsorption and desorption, even with the corrected surface flux, therefore does not ensure conservation. Here we derive the remaining mass defect, construct positive periodic states that exhibit it, and obtain a conservative bulk flux from the moving-boundary balance. We then compare both formulations using the same hydrodynamic equations, parameters and initial condition. Throughout the paper we distinguish conservation of the reduced inventory from higher-order compatibility with the full curved-interface problem.

\section{Formulation and reduced inventories}
\subsection{Geometry, notation and scaling}
Consider a two-dimensional, isothermal, incompressible film on a rigid plane inclined at angle $\theta$. The liquid occupies $0\leq z\leq h(x,t)$, with $x$ directed down the plane. The wall is impermeable and satisfies no slip. All fields are periodic on $0\leq x\leq L$. Let $C(x,z,t)$ and $\Gamma(x,t)$ denote the bulk and surface concentrations. Define
\begin{equation}
 q=\int_0^h u\,\dd z,\qquad
 \varphi=C(x,h(x,t),t),\qquad
 \chi=\int_0^h(C-\varphi)\,\dd z,\qquad B=\chi+h\varphi.
 \label{eq:definitions}
\end{equation}
Thus $B$ is the bulk surfactant inventory per unit streamwise length.

Except in Appendix~\ref{app:scaling}, we use the coordinates of Ref.~\cite{Mukhopadhyay2026PRF}, in which the aspect ratio has been absorbed into the independent variables. With reference thickness $\hat h_N$, speed $\hat V$, kinematic viscosity $\nu$ and surface packing concentration $\hat\Gamma_\infty$, the relevant groups are
\begin{equation}
 \begin{gathered}
 \Rey=\frac{\hat V\hat h_N}{\nu},\qquad
 \Pe_b=\frac{\hat V\hat h_N}{D_b},\qquad
 \Pe_s=\frac{\hat V\hat h_N}{D_s},\\
 k_s=\frac{k_2\hat h_N}{\hat V},\qquad
 \kappa=\frac{k_1}{k_2\hat h_N},\qquad M=\Rey\Mn.
 \end{gathered}
 \label{eq:groups}
\end{equation}
Here $D_b$ and $D_s$ are diffusivities, while $k_1$ and $k_2$ specify adsorption and desorption. Bulk concentration is scaled with $\hat\Gamma_\infty/\hat h_N$ and surface concentration with $\hat\Gamma_\infty$, so the bulk and surface inventories have compatible units. For a linear surface-tension law, $\Mn=\sigma_\Gamma\hat\Gamma_\infty/(\rho\hat V^2\hat h_N)$, where $\sigma_\Gamma>0$ is the magnitude of the dimensional surface-tension derivative. As in Ref.~\cite{Mukhopadhyay2026PRF}, $\sigma_0$ denotes the tension at equilibrium concentration: $\sigma/\sigma_0=1-\Ca M(\Gamma-\Gamma_e)$. This convention is also used in the numerical admissibility check.

With $s=(1+h_x^2)^{1/2}$, distinguish the exact geometric inventory from the projected inventory used by the reduced equations:
\begin{equation}
 \mass_{\rm geom}=\int_0^L(B+\Gamma s)\,\dd x,
 \qquad \mass=\int_0^L(B+\Gamma)\,\dd x.
 \label{eq:inventories}
\end{equation}
The present audit concerns $\mass$. Conservation of $\mass_{\rm geom}$ requires the surface metric in both the evolution equation and the diagnostic, as discussed in Sec.~\ref{sec:geometry}.

\subsection{Profiles and transport equations}
For the no-slip case, set $y=z/h$ and write the retained profiles as
\begin{align}
 u&=\frac{3q}{h}\left(y-\frac{y^2}{2}\right)+\frac{Mh}{4}\Gamma_x(2y-3y^2), \label{eq:u-profile}\\
 C&=\varphi+\frac{3\chi}{2h}(1-y^2). \label{eq:c-profile}
\end{align}
The Marangoni velocity correction has zero depth integral, preserving the definition of $q$. The leading adsorption-desorption relation is
\begin{equation}
 R=k_s[\kappa(1-\Gamma)\varphi-\Gamma]=\frac{3\chi}{\Pe_bh^2}.
 \label{eq:kinetics}
\end{equation}
Its limitations at finite slope are discussed in Sec.~\ref{sec:geometry}.

Integration of $uC$ and evaluation of the surface velocity give
\begin{align}
 F_b^a&=q\varphi+\frac{33q\chi}{40h}+\frac{3M}{80}h\chi\Gamma_x, \label{eq:adv-flux}\\
 F_s&=\frac{3q\Gamma}{2h}-\frac{\Gamma_x}{\Pe_s}-\frac{M}{4}h\Gamma\Gamma_x. \label{eq:surface-flux}
\end{align}
The superscript $a$ denotes bulk advection. The surface flux already includes surface diffusion and the corrected Marangoni term.

For $\beta=0$, Eqs.~(56)-(57) of Ref.~\cite{Mukhopadhyay2026PRF} become
\begin{align}
 B_t+\partial_xF_b^a&=\frac{1}{\Pe_b}\left(\chi_{xx}+h\varphi_{xx}-\frac{3\chi h_x^2}{h^2}\right)-R, \label{eq:reference-bulk}\\
 \Gamma_t+\partial_xF_s&=R. \label{eq:reference-surface}
\end{align}
The same bulk diffusion bracket appears in Eq.~(34) of Ref.~\cite{Mukhopadhyay2026POF}, with prefactor $\varepsilon/\Pe_b$ in its explicit long-wave coordinates. Its surface equation~(35) also uses the corrected Marangoni coefficient. Appendix~\ref{app:scaling} relates the two scalings.

\section{Mass balance of the reference model}
\subsection{The mass defect}
Adding Eqs.~\eqref{eq:reference-bulk} and \eqref{eq:reference-surface} cancels $R$. Periodicity removes the flux derivatives and $\chi_{xx}$. Integrating $h\varphi_{xx}$ by parts then yields
\begin{equation}
 \frac{\dd\mass}{\dd t}=-\frac{1}{\Pe_b}\int_0^L
 \left(h_x\varphi_x+\frac{3\chi h_x^2}{h^2}\right)\dd x.
 \label{eq:defect}
\end{equation}
This expression need not vanish and can have either sign. It identifies a defect in the differential equations themselves, independently of the grid or time-integration method.

In Sec.~V of Ref.~\cite{Mukhopadhyay2026POF}, the exchange terms cancel and the Marangoni terms are correctly identified as fluxes. The remaining bulk-diffusion contribution has the integral in Eq.~\eqref{eq:defect}. The exact conservation statement in Eq.~(41) of that paper therefore does not hold for general periodic solutions. The same residual occurs in Ref.~\cite{Mukhopadhyay2026PRF}.

\subsection{Positive periodic counterexamples}
Let
\begin{equation}
 \begin{gathered}
 k=2\pi/L,\qquad h=1+a\cos(kx),\qquad
 \varphi=\varphi_0+d\cos(kx+\alpha),\\
 \chi=0,\qquad \Gamma=\frac{\kappa\varphi}{1+\kappa\varphi},\qquad q=\frac23,
 \end{gathered}
 \label{eq:test-state}
\end{equation}
where $0<a<1$, $|d|<\varphi_0$ and $\kappa>0$. These fields satisfy $h>0$, $C=\varphi>0$, $0<\Gamma<1$ and the reduced kinetic relation with $R=0$. They are admissible instantaneous states of the reduced system; the test does not require compatibility with the full finite-slope boundary conditions. Equation~\eqref{eq:defect} gives
\begin{equation}
 \left.\frac{\dd\mass}{\dd t}\right|_{\rm ref}
 =-\frac{adk^2L}{2\Pe_b}\cos\alpha.
 \label{eq:first-rate}
\end{equation}
For $L=500$, $\Pe_b=700$, $\varphi_0=1/90$, $\kappa=10$, $a=0.05$, $d=0.001$ and $\alpha=0$, the rate is $-2.81988697174\times10^{-9}$. Reversing $d$ produces mass growth. A phase difference $\alpha=\pi/2$ makes this example's rate vanish, showing why a single zero-drift test cannot establish conservation.

To isolate the second term in Eq.~\eqref{eq:defect}, take $\varphi=\varphi_0$ and $\chi=\chi_0>0$ constant, and set
\begin{equation}
 \Gamma=\frac{\kappa\varphi_0-3\chi_0/(\Pe_bk_sh^2)}{1+\kappa\varphi_0}.
 \label{eq:second-state}
\end{equation}
The explicit bound $0<\chi_0<\Pe_b k_s\kappa\varphi_0(1-a)^2/3$ ensures $0<\Gamma<1$; the reconstructed bulk profile is then positive and the kinetic relation is satisfied. Then
\begin{equation}
 \frac{\dd\mass}{\dd t}=-\frac{3\chi_0k^2L}{\Pe_b}
 \left(\frac{1}{\sqrt{1-a^2}}-1\right)<0.
 \label{eq:second-rate}
\end{equation}
With the same parameters, $k_s=0.5$ and $\chi_0=0.001$, this gives $-4.23777795\times10^{-10}$.

\subsection{Magnitude, sign and linearization}
For any smooth solution with $h>0$, Eq.~\eqref{eq:defect} implies
\begin{equation}
 |\mass(t)-\mass(0)|\leq\frac{1}{\Pe_b}\int_0^t
 \left[\|h_x\|_2\|\varphi_x\|_2+3\left\|\frac{\chi}{h^2}\right\|_\infty\|h_x\|_2^2\right]\dd\tau.
 \label{eq:bound}
\end{equation}
The norms are over one period. The imbalance depends on the thickness and concentration gradients, the bulk departure $\chi$, and the minimum thickness. The factor $\Pe_b^{-1}$ applies to a specified solution; changing $\Pe_b$ also changes the evolving fields.

Expand about uniform adsorption equilibrium:
\begin{equation}
 h=h_0+\delta h_1,\quad\varphi=\varphi_0+\delta\varphi_1,
 \quad\chi=\delta\chi_1,\quad\Gamma=\Gamma_0+\delta\Gamma_1.
 \label{eq:linear-expansion}
\end{equation}
The first term in the correction is $\delta^2h_{1x}\varphi_{1x}/\Pe_b$, while the term proportional to $\chi h_x^2$ begins at cubic order. The first variation is zero. Both reduced models therefore have the same linearization about uniform equilibrium, although their nonlinear evolution differs.

\begin{figure}
\centering\includegraphics[width=.95\textwidth]{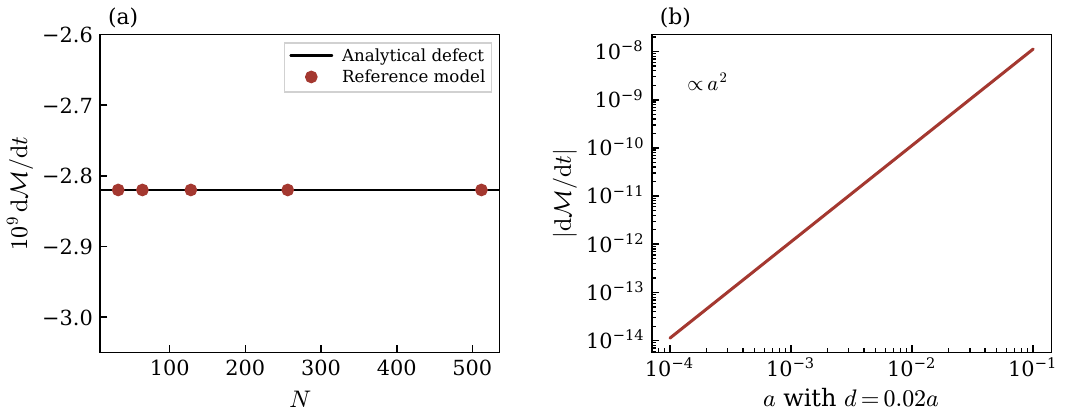}
\caption{Instantaneous mass-balance test. (a) Fourier evaluation of the complete reference transport pair for $a=0.05$, $d=0.001$ and $\alpha=0$ reproduces Eq.~\eqref{eq:first-rate} under grid refinement. Isolating bulk diffusion gives the same total rate. (b) With $d=0.02a$, the analytical defect scales quadratically with amplitude.}
\label{fig:instantaneous}
\end{figure}

\section{Conservative bulk-surface reduction}
\subsection{Integrate the bulk balance before closing the flux}
The bulk concentration satisfies
\begin{equation}
 C_t+\partial_x(uC)+\partial_z(wC)=\frac{C_{xx}+C_{zz}}{\Pe_b}.
 \label{eq:bulk-pde}
\end{equation}
At the wall, $w=C_z=0$; at the free surface, $h_t+u(h)h_x=w(h)$. Define
\begin{equation}
 F=\int_0^h uC\,\dd z,\qquad Q=\int_0^h C_x\,\dd z.
 \label{eq:exact-fluxes}
\end{equation}
Integrating Eq.~\eqref{eq:bulk-pde} and using the kinematic condition gives
\begin{equation}
 B_t+F_x=\frac{1}{\Pe_b}\left[\int_0^h C_{xx}\,\dd z+C_z(h)\right].
 \label{eq:integrated-bulk}
\end{equation}
Derivatives inside the depth integrals are taken at fixed $z$.

Let $j_n$ be the adsorption flux into the surface per unit actual interface length, and let $J=sj_n$ be its projected counterpart. With outward unit normal $\bm n=(-h_x,1)/s$, Fick's law gives
\begin{equation}
 J=\frac{h_xC_x(h)-C_z(h)}{\Pe_b},\qquad
 C_z(h)=h_xC_x(h)-\Pe_bJ.
 \label{eq:normal-flux}
\end{equation}
Leibniz' rule also gives
\begin{align}
 Q&=B_x-\varphi h_x=\chi_x+h\varphi_x, \label{eq:Q}\\
 Q_x&=\int_0^h C_{xx}\,\dd z+h_xC_x(h). \label{eq:Qx}
\end{align}
Substitution into Eq.~\eqref{eq:integrated-bulk} yields the exact identity
\begin{equation}
 B_t+F_x=\frac{1}{\Pe_b}\partial_x(\chi_x+h\varphi_x)-J.
 \label{eq:exact-depth}
\end{equation}
No polynomial concentration profile was needed for this step.

The quadratic profile~\eqref{eq:c-profile} makes the source of the reference-model defect explicit:
\begin{equation}
 C_z(h)=-\frac{3\chi}{h^2},\qquad
 C_x(h)=\varphi_x+\frac{3\chi h_x}{h^2},
 \label{eq:surface-gradients}
\end{equation}
and hence
\begin{equation}
 \int_0^hC_{xx}\,\dd z=\chi_{xx}+h\varphi_{xx}-\frac{3\chi h_x^2}{h^2}.
 \label{eq:diffusion-integral}
\end{equation}
This is precisely the diffusion bracket in Eq.~\eqref{eq:reference-bulk}. It accounts for the depth integral of $C_{xx}$, but the normal-flux condition contributes the additional boundary term $h_xC_x(h)/\Pe_b$. Omitting that term identifies exchange with vertical diffusion while retaining streamwise diffusion over a sloping film.

\subsection{The corrected transport system}
Retain the projected surface balance, the kinetic relation~\eqref{eq:kinetics} and the advective fluxes~\eqref{eq:adv-flux}-\eqref{eq:surface-flux}. Approximating $J$ by the same $R$ in both reservoirs gives
\begin{align}
 B_t+\partial_xF_b&=-R,\qquad F_b=F_b^a-\frac{\chi_x+h\varphi_x}{\Pe_b}, \label{eq:corrected-bulk}\\
 \Gamma_t+\partial_xF_s&=R. \label{eq:corrected-surface}
\end{align}
The addition to the right-hand side of Eq.~\eqref{eq:reference-bulk} is
\begin{equation}
 \Delta_b=\frac{1}{\Pe_b}\left(h_x\varphi_x+\frac{3\chi h_x^2}{h^2}\right).
 \label{eq:correction}
\end{equation}
Adding the corrected equations gives the local conservation law
\begin{equation}
 \partial_t(B+\Gamma)+\partial_x(F_b+F_s)=0.
 \label{eq:local-conservation}
\end{equation}
For periodic boundaries, $\dd\mass/\dd t=0$. For nonperiodic boundaries, the inventory change is determined by the net boundary flux. Neither statement requires an adjustment of the spatial mean. The hydrodynamic equations remain unchanged; the complete model is given in Appendix~\ref{app:model}.

\subsection{Storage variables and discrete conservation}
For time integration, evolving $(h,q,B,\Gamma)$ avoids differentiating the nonlinear relation between $\chi$, $h$, $\varphi$ and $\Gamma$. With $a_s=\Pe_bk_sh^2/3$, the concentration variables follow algebraically:
\begin{equation}
 \varphi=\frac{B+a_s\Gamma}{h+a_s\kappa(1-\Gamma)},\qquad \chi=B-h\varphi.
 \label{eq:reconstruction}
\end{equation}
The denominator is positive for $h>0$, $0\leq\Gamma<1$ and positive kinetic parameters. Nonnegative bulk concentration is a separate requirement and is monitored during each computation.

Let $D$ be a periodic discrete derivative with $\bm1^TD=0$. Using the same discrete exchange vector with opposite signs gives
\begin{equation}
 \frac{\dd}{\dd t}\left[\Delta x\,\bm1^T(B+\Gamma)\right]
 =-\Delta x\,\bm1^TD(F_b+F_s)=0.
 \label{eq:discrete}
\end{equation}
For Fourier differentiation, this is the zero-wavenumber identity. It checks the discrete inventory balance independently of how accurately the remaining modes resolve the waveform.

\subsection{Reduced conservation and finite-slope compatibility}
\label{sec:geometry}
The exact graph form of the material-surface balance is
\begin{equation}
 (\Gamma s)_t+\partial_x(\Gamma sU)=\frac{1}{\Pe_s}\partial_x\left(\frac{\Gamma_x}{s}\right)+J,
 \qquad U=u(x,h(x,t),t).
 \label{eq:exact-surface}
\end{equation}
Here $U$ is the horizontal velocity of a material point on the graph. The evolving measure $s\,\dd x$ accounts for surface deformation. Together with Eq.~\eqref{eq:exact-depth}, this equation conserves $\mass_{\rm geom}$.

The retained quadratic profile and leading kinetic relation do not jointly satisfy the full finite-slope normal-flux condition. Substituting $j_n=k_s[\kappa(1-\Gamma)\varphi-\Gamma]$ into Eq.~\eqref{eq:normal-flux} requires
\begin{equation}
 \frac{3\chi}{h^2}(1+h_x^2)+h_x\varphi_x
 =\Pe_b s k_s[\kappa(1-\Gamma)\varphi-\Gamma].
 \label{eq:compatibility}
\end{equation}
The leading relation~\eqref{eq:kinetics} omits these slope terms. Extending the geometric order therefore requires a corresponding revision of the concentration profile or its closure and of the surface balance.

Equations~\eqref{eq:corrected-bulk}-\eqref{eq:corrected-surface} conserve the projected inventory exactly within the stated reduction. The flux correction does not, by itself, provide a complete higher-order approximation to the full curved-interface problem.

\section{Numerical comparison}
\subsection{Parameters, initial condition and numerical method}
Both models are integrated with identical hydrodynamics, parameters and initial data. The benchmark uses the no-slip case, $\beta=0$, and the parameters reported for Fig.~10 of Ref.~\cite{Mukhopadhyay2026PRF}:
\begin{equation}
 \begin{gathered}
 \Rey=30,\quad\theta=4^\circ,\quad\Ka=3905,\quad\Mn=0.2,\quad\Gamma_e=0.1,\\
 \kappa=10,\quad\Pe_b=\Pe_s=700,\quad k_s=0.5,\quad L=500,\quad t_f=4500.
 \end{gathered}
 \label{eq:benchmark}
\end{equation}
The velocity scale is the Nusselt surface speed, giving $G=\Rey/\mathrm{Fr}^2=2$ and $q_e=G/3=2/3$. With the Kapitza convention of Ref.~\cite{Mukhopadhyay2026PRF},
\begin{equation}
 \Ca=\frac{\Rey^{2/3}}{G^{1/3}\Ka}=0.0019623801,\qquad
 M=6,\qquad\varphi_e=\frac{\Gamma_e}{\kappa(1-\Gamma_e)}=\frac1{90}.
 \label{eq:capillarity}
\end{equation}
These choices define a dimensionless comparison between the two formulations.

The initial perturbation follows the most unstable eigenvector $\bm v$ at the fundamental wavenumber of their common linearized operator:
\begin{equation}
 \bm u(x,0)=\bm u_e+0.01\operatorname{Re}\{\bm v e^{2\pi ix/L}\},\quad
 \bm u=(h,q,B,\Gamma)^T,\quad \bm u_e=(1,2/3,1/90,0.1)^T,
 \label{eq:initial}
\end{equation}
with $v_h=1$. The growth rate is $8.10212\times10^{-4}$ and the phase speed is $1.97658$. The results below come from an independent implementation of the equations, rather than a reanalysis of the data behind the earlier traveling-wave figure.

Spatial derivatives are evaluated by Fourier collocation with $N=512$. Nonlinear expressions are evaluated on a grid padded to $2N$ and projected back; the coarse-grid Nyquist mode is excluded consistently. Padding reduces aliasing, but cannot remove it exactly for the rational concentration closure. The state is stored as a Fourier perturbation about $\bm u_e$.

The fourth-order exponential Runge-Kutta formula of Cox and Matthews~\cite{CoxMatthews2002} advances the system. The common uniform-state linear operator is treated through $4\times4$ matrix functions, evaluated with augmented block exponentials. Adaptive step doubling accepts two half steps when the maximum componentwise error estimate, divided by $15$, is below $10^{-9}+10^{-6}\max(|u_{\rm old}|,|u_{\rm fine}|)$. The maximum nominal step is $0.1$; the accepted update therefore uses two steps no larger than $0.05$. Smaller steps are dyadic subdivisions of the nominal ceiling. The mixed derivative $\Gamma_{tx}$ is evaluated from the current surface right-hand side, as in Appendix~\ref{app:model}.

At zero wavenumber, the coefficient matrices satisfy the analytical linear inventory identities. This is a property of the time-integration coefficients, not a correction to the solution or its right-hand side. The reference model's nonlinear zero-mode source is retained. No concentration clipping or mass renormalization is used. Film thickness, the reconstructed bulk concentration, surface saturation and equilibrium-referenced surface tension remain in their admissible ranges at all monitored points of the padded grid at every accepted step. These checks do not constitute a positivity theorem for the differential equations. The computation notebook provides the solver and verification calculations; a separate plotting notebook reads the supplied numerical tables.

\subsection{Inventory balance and wave profiles}
\begin{figure}
\centering\includegraphics[width=\textwidth]{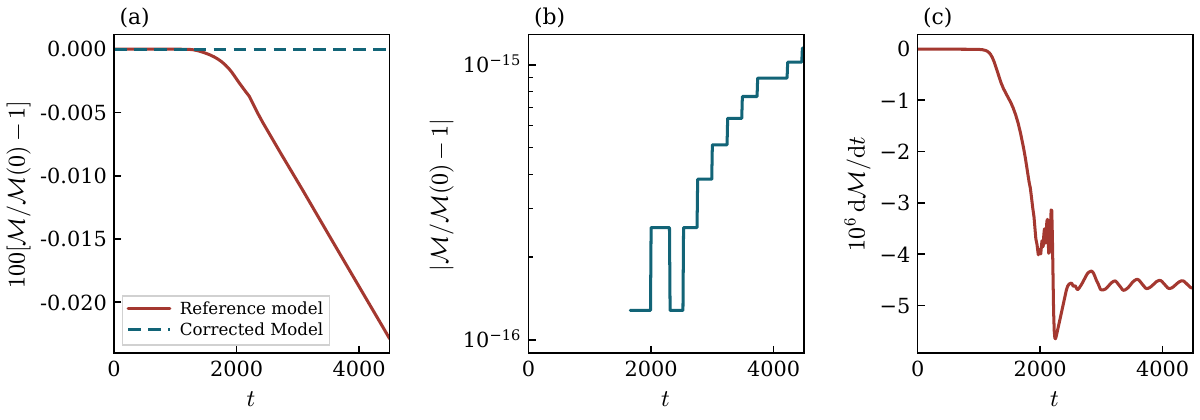}
\caption{Mass balance in the newly computed benchmark. (a) Signed relative inventory change in percent. (b) Absolute relative inventory change for the corrected model; exact-zero samples are omitted from the logarithmic plot. (c) The rate from Eq.~\eqref{eq:defect}, evaluated along the reference trajectory. Both runs use Eqs.~\eqref{eq:benchmark} and \eqref{eq:initial}, without mass adjustment.}
\label{fig:mass}
\end{figure}
Define the signed relative inventory change
\begin{equation}
 E_{\mass}(t)=\frac{\mass(t)-\mass(0)}{\mass(0)}.
 \label{eq:relative-mass}
\end{equation}
Both runs start with $\mass(0)=55.5555555556$ and reach $t_f=4500$. The newly computed endpoint values are
\begin{equation}
 E_{\mass}^{\rm ref}=\ReferenceMassChange,\qquad
 E_{\mass}^{\rm c}=\CorrectedMassChange.
 \label{eq:mass-result}
\end{equation}
The reference model loses approximately $\ReferenceLossPercent\%$ of its initial inventory. The largest absolute relative inventory change among the $451$ saved samples is also $1.15108\times10^{-15}$ in the corrected run. Its inventory balance is therefore satisfied to floating-point accuracy. The precise residual at that level depends on the representation of the state and the numerical implementation; it is not a measure of the overall solution error.

At the reference endpoint, Eq.~\eqref{eq:defect} gives a mass rate of $\ReferenceFinalRate$. It agrees with the zero Fourier mode of the complete transport right-hand side to within $\RateIdentityError$ over the saved trajectory. For the corrected equations, the total rate is zero up to floating-point evaluation. This agreement identifies the bulk residual as the source of the drift.

\begin{figure}
\centering\includegraphics[width=\textwidth]{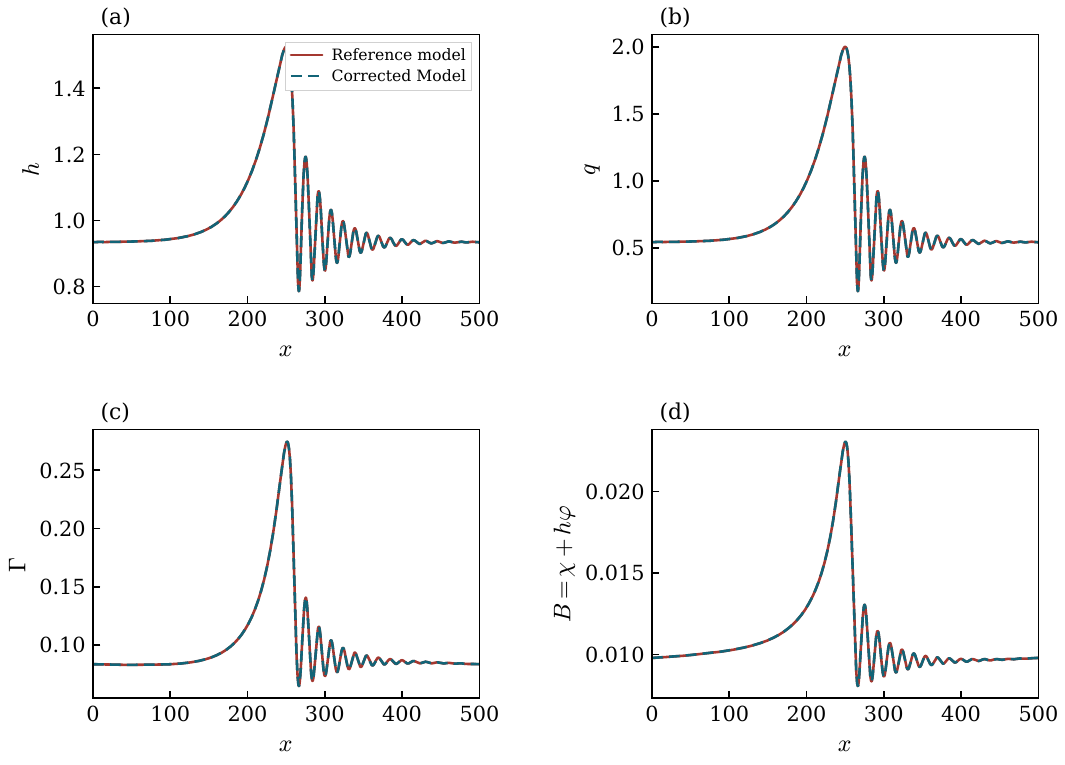}
\caption{Phase-aligned profiles at $t=4500$: (a) film thickness, (b) flow rate, (c) surface concentration and (d) bulk inventory. The height maxima are placed at $x=250$. The close agreement shows that removing the mass defect changes this benchmark's waveform only slightly.}
\label{fig:profiles}
\end{figure}

\begin{table}
\centering\small
\caption{Diagnostics at $t=4500$. Extrema are evaluated from the numerical Fourier representations on $32768$ points. This interpolation reduces sampling bias without increasing the simulation resolution.}
\label{tab:endpoint}
\begin{tabular}{@{}lrr@{}}\toprule
Diagnostic & Reference model & Corrected model\\\midrule
Minimum $h$ & 0.78577611 & 0.78578015\\
Maximum $h$ & 1.52565449 & 1.52563997\\
Maximum $q$ & 2.00177542 & 2.00173249\\
Maximum $\Gamma$ & 0.27439575 & 0.27444423\\
Minimum $\varphi$ & 0.00694174 & 0.00694336\\
Minimum wall concentration & 0.00372036 & 0.00372517\\
Fitted translation speed & 2.46519355 & 2.46518324\\
Signed relative mass change & $-2.28832\times10^{-4}$ & $-1.15108\times10^{-15}$\\
\bottomrule\end{tabular}

\end{table}
Figure~\ref{fig:profiles} compares the terminal profiles. A single translation of all four fields in each solution places its principal height maximum at $x=250$. This removes the relative phase without changing either inventory. Both models develop a main hump and a train of capillary ripples. The maximum differences between the aligned profiles are
\begin{equation}
 \begin{aligned}
 \|h^{\rm c}-h^{\rm r}\|_\infty&=\DifferenceH,&
 \|q^{\rm c}-q^{\rm r}\|_\infty&=\DifferenceQ,\\
 \|B^{\rm c}-B^{\rm r}\|_\infty&=\DifferenceB,&
 \|\Gamma^{\rm c}-\Gamma^{\rm r}\|_\infty&=\DifferenceGamma.
 \end{aligned}
 \label{eq:profile-differences}
\end{equation}
Superscripts $\rm r$ and $\rm c$ denote the reference and corrected computations.

The translation speed $c$ is fitted by minimizing the squared norm of $\bm u_t+c\bm u_x$, after scaling the fields by $(1,2/3,1/90,0.1)$. The residual, normalized by the norm of the scaled $\bm u_t$, is approximately $\TranslationResidual$ in both runs. The profiles are nearly translating waves at this time, but still undergo slight deformation; they are not asserted to be exact permanent waves.

\begin{table}
\centering\small
\caption{Maximum absolute changes in terminal fields in late-time refinement tests. ``Time'' compares the two step ceilings at $N=512$; ``space'' compares $N=512$ and $1024$ at the smaller ceiling. All continuations cover $4400\leq t\leq4500$.}
\label{tab:refinement}
\begin{tabular}{@{}llrrrr@{}}\toprule
Model & Test & $h$ & $q$ & $B$ & $\Gamma$\\\midrule
Reference & Time & $5.53\times10^{-8}$ & $2.31\times10^{-7}$ & $6.56\times10^{-10}$ & $1.02\times10^{-8}$\\
Reference & Space & $1.22\times10^{-7}$ & $3.01\times10^{-7}$ & $1.58\times10^{-9}$ & $1.58\times10^{-8}$\\
Corrected & Time & $5.53\times10^{-8}$ & $2.30\times10^{-7}$ & $6.56\times10^{-10}$ & $1.02\times10^{-8}$\\
Corrected & Space & $1.22\times10^{-7}$ & $3.01\times10^{-7}$ & $1.58\times10^{-9}$ & $1.58\times10^{-8}$\\
\bottomrule\end{tabular}

\end{table}

\subsection{Numerical verification}
The complete transport pair is evaluated on the analytical states with $N=32,64,128,256,512$. Both signs of the first example, its quadrature case and the constant-$\chi$ example reproduce Eqs.~\eqref{eq:first-rate} and \eqref{eq:second-rate}. The maximum absolute corrected rate is $\InstantaneousConservativeRate$. The reference rates persist under refinement. Independent symbolic calculations verify the profile integrals, the fixed-$z$ derivatives, the boundary contribution and the vanishing first variation of the correction.

Three continuations are performed for each model from its own state saved at $t=4400$: the baseline uses $N=512$ and $\Delta t_{\max}=0.1$; the temporal check uses $N=512$ and $\Delta t_{\max}=0.025$; and the spatial check uses $N=1024$ and $\Delta t_{\max}=0.025$. The same Fourier state starts all three continuations for a given model, with zero padding for the larger grid. The final fields are compared on a common dense grid without phase alignment.

The changes in Table~\ref{tab:refinement} are smaller than the model differences in Eq.~\eqref{eq:profile-differences}. They support the comparison over the tested late-time interval. Because the continuations begin at $t=4400$, they do not bound errors accumulated earlier in the trajectory. The analytical conservation identity and instantaneous counterexamples do not depend on this limitation.

\section{Discussion} The surface and bulk inconsistencies arise from different parts of the reduced transport model. Replacing the surface coefficient $5/4$ by $1/4$ removes the Marangoni residual identified in Refs.~\cite{Mukhopadhyay2026PRF,Mukhopadhyay2026POF}, but it does not affect the bulk contribution examined here. In the isothermal limit of Ref.~\cite{DAlessio2020}, written in the present concentration normalization, the rate of change of the total inventory contains both terms: \begin{equation} \frac{\dd\mass}{\dd t}=M\int_0^L\Gamma\Gamma_xh_x\,\dd x -\frac{1}{\Pe_b}\int_0^L\left(h_x\varphi_x+\frac{3\chi h_x^2}{h^2}\right)\dd x. \label{eq:both-residuals} \end{equation} The first integral is removed by the surface correction, whereas Eq.~\eqref{eq:correction} removes the second. For a closed periodic system, conservation requires the entire right-hand side to vanish identically. The present comparison is limited to the isothermal transport closure; additional thermal and variable-density effects considered in Refs.~\cite{Pascal2019,DAlessio2020} are beyond the scope of this work. For the travelling-wave benchmark considered here, correcting the bulk transport equation changes the waveform only weakly. The principal hump, the capillary ripples and the propagation speed remain very close to those of the reference computation. Nevertheless, the reference model does not satisfy the exact conservation law implied by the reduced bulk--surface transport problem. The analytical calculation and its persistence under numerical refinement show that the observed drift is not a numerical artifact but a property of the governing equations themselves. The correction restores conservation of the reduced total inventory and makes that conservation law an intrinsic feature of the nonlinear model. Although the drift is small in this particular benchmark, exact conservation should not depend on parameter choice, simulation time or the eventual inclusion of additional physical effects. The corrected bulk flux follows directly from the depth-integrated advection-diffusion equation together with the normal-flux condition at the free surface. In practice, the safest route is to integrate the bulk balance first and only then introduce an approximate concentration profile. The same result can be obtained by substituting the profile earlier in the derivation, provided that the boundary terms generated by Leibniz' rule are retained. The argument is therefore not tied to the specific quadratic approximation used here. More broadly, it provides a straightforward way of maintaining bulk-surface mass balance when deriving or extending reduced transport closures. Finally, conservation alone is not sufficient to establish the overall validity of a reduced model. Comparison with the parent free-boundary problem remains necessary to assess the accuracy of the retained concentration profile and the range of applicability of the closure. The present modification establishes exact conservation of the projected bulk-surface inventory within the existing reduced description. Questions of higher-order compatibility with the full curved-interface geometry require a corresponding extension of the asymptotic closure and are left for future work.

\section{Conclusions} We have shown that the weighted-residual models of Refs.~\cite{Mukhopadhyay2026PRF,Mukhopadhyay2026POF} retain a residual bulk-diffusion contribution even after correction of the surface Marangoni term. Its effect on the total surfactant inventory is given by Eq.~\eqref{eq:defect}. Explicit positive periodic concentration fields demonstrate that the resulting mass drift need not vanish, and Fourier calculations confirm that the defect persists under numerical refinement. Revisiting the depth-integrated transport balance shows that a contribution associated with the streamwise concentration gradient in the normal diffusive flux has been omitted. Retaining this term leads to the streamwise diffusive flux $-(\chi_x+h\varphi_x)/\Pe_b$. Together with equal and opposite adsorption--desorption terms in the bulk and surface equations, it restores local conservation of the reduced bulk--surface transport system. The hydrodynamic equations, the leading adsorption--desorption closure, and the linearization about uniform equilibrium are unchanged. For the independent no-slip benchmark considered here, the reference model loses approximately $\ReferenceLossPercent\%$ of its initial surfactant inventory by $t=4500$, whereas the corrected model preserves the inventory to floating-point accuracy. The principal wave hump and capillary ripple structure remain almost unchanged. Therefore, this benchmark illustrates that agreement with traveling-wave calculations does not by itself guarantee that a reduced model respects the conservation laws inherited from the underlying transport problem. In the corrected formulation, the total projected surfactant inventory is conserved for every smooth periodic solution. The present work addresses conservation within the reduced description. Assessing the accuracy of the concentration profile and the range of validity of the closure still requires comparison with the parent free-boundary problem. Likewise, higher-order compatibility with the full curved-interface geometry will require a corresponding extension of the asymptotic closure.

\section*{Data availability}
The numerical data and verification scripts for this study are available upon request.

\section*{Declaration of Competing Interests}
The authors declare that they have no known competing financial interests or personal relationships that could have appeared to influence the work reported in this paper.

\appendix
\section{Explicit long-wave factors and source correspondence}
\label{app:scaling}
\renewcommand{\theequation}{A\arabic{equation}}\setcounter{equation}{0}
Let $X=\varepsilon x$, $\tau=\varepsilon t$ and $w=\varepsilon W$. Then $\partial_x=\varepsilon\partial_X$ and $\partial_t=\varepsilon\partial_\tau$. The corrected pair becomes
\begin{align}
 B_\tau+\partial_X\left[F_b^{a,\varepsilon}-\frac{\varepsilon}{\Pe_b}(\chi_X+h\varphi_X)\right]
 &=-\frac{3\chi}{\varepsilon\Pe_bh^2},\label{eq:explicit-bulk}\\
 \Gamma_\tau+\partial_X\left[\frac{3q\Gamma}{2h}-\frac{\varepsilon}{\Pe_s}\Gamma_X-\frac{\varepsilon M}{4}h\Gamma\Gamma_X\right]
 &=\frac{3\chi}{\varepsilon\Pe_bh^2},\label{eq:explicit-surface}
\end{align}
where $F_b^{a,\varepsilon}=q\varphi+33q\chi/(40h)+3\varepsilon Mh\chi\Gamma_X/80$. The addition to Eq.~(34) of Ref.~\cite{Mukhopadhyay2026POF} is
\begin{equation}
 \Delta_b^\varepsilon=\frac{\varepsilon}{\Pe_b}\left(h_X\varphi_X+\frac{3\chi h_X^2}{h^2}\right),
\end{equation}
with $M=\Rey\mathrm{Mr}$ in that paper's notation. It has the same explicit long-wave prefactor as the streamwise bulk diffusion already retained.

\begin{center}\small
\begin{tabularx}{\textwidth}{@{}p{.33\textwidth}X@{}}\toprule
Source & Relation to the present formulation\\\midrule
Ref.~\cite{Mukhopadhyay2026PRF}, Eqs.~(56)-(57) & At $\beta=0$, these give Eqs.~\eqref{eq:reference-bulk}-\eqref{eq:reference-surface}.\\[4pt]
Ref.~\cite{Mukhopadhyay2026POF}, Eqs.~(34)-(35) & The same transport pair with explicit $\varepsilon$ factors.\\[4pt]
Ref.~\cite{DAlessio2020}, Eqs.~(4.16)-(4.17) & In the isothermal limit, $\varphi=C_D$ and $\chi=(\Pe_bk_s/3)\chi_D$. The bulk bracket agrees, while the earlier surface residual remains.\\[4pt]
Ref.~\cite{Pascal2019}, Eqs.~(53)-(54) & The surface equation contains the $5/4$ coefficient. The bulk profile and definition of $\chi$ differ, so Eq.~(54) is not directly identified with Eq.~\eqref{eq:reference-bulk}.\\\bottomrule
\end{tabularx}
\end{center}
The parameter called $\Ka$ in Ref.~\cite{Mukhopadhyay2026POF} is $\varepsilon^2\Rey\mathrm{We}$. Ref.~\cite{Mukhopadhyay2026PRF} instead uses the Kapitza number $\Ka=\sigma_0/[\rho\nu^{4/3}(g\sin\theta)^{1/3}]$, which is the convention used in the numerical comparison.

In explicit long-wave coordinates, $s_\varepsilon=(1+\varepsilon^2h_X^2)^{1/2}$ and Eq.~\eqref{eq:compatibility} becomes
\begin{equation}
 \frac{3\chi}{h^2}(1+\varepsilon^2h_X^2)+\varepsilon^2h_X\varphi_X
 =\Pe_bs_\varepsilon k_s[\kappa(1-\Gamma)\varphi-\Gamma].
\end{equation}
The surface metric and normal diffusion terms must be ordered together in any further asymptotic extension.

\section{The corrected nonlinear weighted-residual model}
\label{app:model}
\renewcommand{\theequation}{B\arabic{equation}}\setcounter{equation}{0}
For an isothermal film on a rigid, impermeable, no-slip wall, write $M=\Rey\Mn$, $G=\Rey/\mathrm{Fr}^2$ and $\Ca=(\Rey\mathrm{We})^{-1}$. Here $\mathrm{Fr}=\hat V/(g\hat h_N\sin\theta)^{1/2}$ and $\mathrm{We}=\sigma_0/(\rho\hat V^2\hat h_N)$, with reference surface tension $\sigma_0$ and density $\rho$. The model is
\begin{equation}
 h_t+q_x=0,\label{eq:B1}
\end{equation}
\begin{equation}
\begin{aligned}
q_t+\frac{17q}{7h}q_x-\frac{9q^2}{7h^2}h_x
={}&\frac{5}{2\Rey}\left(\frac{Gh}{3}-\frac{q}{h^2}-\frac M2\Gamma_x\right)
 +\frac{5h}{6\Rey}\left(\frac{h_{xxx}}{\Ca}-G\cot\theta\,h_x\right)\\
&+\frac1{\Rey}\left(\frac{4qh_x^2}{h^2}-\frac{9q_xh_x}{2h}-\frac{6qh_{xx}}h+\frac92q_{xx}\right)\\
&+M\left(\frac{h^2\Gamma_{tx}}{48}+\frac{15qh\Gamma_{xx}}{224}
 +\frac{19hq_x\Gamma_x}{336}+\frac{5qh_x\Gamma_x}{112}\right),
\end{aligned}\label{eq:B2}
\end{equation}
\begin{equation}
 \partial_t(\chi+h\varphi)+\partial_x\left[q\varphi+\frac{33q\chi}{40h}
 +\frac{3M}{80}h\chi\Gamma_x-\frac{\chi_x+h\varphi_x}{\Pe_b}\right]
 =-\frac{3\chi}{\Pe_bh^2},\label{eq:B3}
\end{equation}
\begin{equation}
 \Gamma_t+\partial_x\left[\frac{3q\Gamma}{2h}-\frac{\Gamma_x}{\Pe_s}-\frac M4h\Gamma\Gamma_x\right]
 =\frac{3\chi}{\Pe_bh^2}.\label{eq:B4}
\end{equation}
Equations~\eqref{eq:B1}-\eqref{eq:B2} are the hydrodynamic equations of Ref.~\cite{Mukhopadhyay2026PRF} at $\beta=0$. Equation~\eqref{eq:B4} retains its corrected surface flux; only the bulk diffusion in Eq.~\eqref{eq:B3} is changed.

The leading adsorption-desorption closure is
\begin{equation}
 \frac{3\chi}{\Pe_bh^2}=k_s[\kappa(1-\Gamma)\varphi-\Gamma],\qquad
 \varphi=\frac{\Gamma+3\chi/(\Pe_bk_sh^2)}{\kappa(1-\Gamma)}.\label{eq:B5}
\end{equation}
When $(h,q,B,\Gamma)$ is evolved, recover
\begin{equation}
 \varphi=\frac{B+a_s\Gamma}{h+a_s\kappa(1-\Gamma)},\quad
 \chi=B-h\varphi,\quad
 C=\varphi+\frac{3\chi}{2h}\left(1-\frac{z^2}{h^2}\right),\quad
 a_s=\frac{\Pe_bk_sh^2}{3}.\label{eq:B6}
\end{equation}
The admissible concentration range requires $0\leq\Gamma<1$, $\varphi\geq0$ and $\varphi+3\chi/(2h)\geq0$. The last two conditions are equivalent to nonnegativity of this quadratic profile throughout the film.

Differentiating the surface equation supplies the mixed derivative in Eq.~\eqref{eq:B2}:
\begin{equation}
 \Gamma_{tx}=-\frac32\partial_{xx}\left(\frac{q\Gamma}{h}\right)
 +\frac{\Gamma_{xxx}}{\Pe_s}+\frac M4\partial_{xx}(h\Gamma\Gamma_x)
 +\frac3{\Pe_b}\partial_x\left(\frac{\chi}{h^2}\right).\label{eq:B7}
\end{equation}
It introduces no additional evolution equation. All terms are determined from the current fields.

The uniform equilibrium is
\begin{equation}
 h=h_0>0,\qquad q=\frac{Gh_0^3}{3},\qquad\Gamma=\Gamma_e,\qquad
 \chi=0,\qquad\varphi=\frac{\Gamma_e}{\kappa(1-\Gamma_e)}.\label{eq:B8}
\end{equation}
For periodic fields and compatible derivatives, the complete flux identity is
\begin{equation}
\begin{aligned}
 &\partial_t(\chi+h\varphi+\Gamma)+\partial_x\mathcal F=0,\\
 &\mathcal F=q\varphi+\frac{33q\chi}{40h}+\frac{3M}{80}h\chi\Gamma_x
 -\frac{\chi_x+h\varphi_x}{\Pe_b}+\frac{3q\Gamma}{2h}
 -\frac{\Gamma_x}{\Pe_s}-\frac M4h\Gamma\Gamma_x.
\end{aligned}\label{eq:B9}
\end{equation}
The model conserves both $\int_0^Lh\,\dd x$ and $\int_0^L(B+\Gamma)\,\dd x$. These are invariants of the stated projected closure; finite-slope compatibility is a separate requirement.

\end{document}